# Talking about responsible quantum: "Awareness is the absolute minimum that … we need to do."

**Dr Tara Roberson**\*, Australian Research Council Centre of Excellence for Engineered Quantum Systems, University of Queensland, Australia

\*corresponding author, t.roberson@uq.edu.au

**Abstract**
Hype over novel quantum technologies has prompted discussion on the likely societal impacts of the sector. Calls to ensure the responsible development of quantum technologies are complicated by a lack of concrete case studies or real-world examples of irresponsible quantum. At this stage, responsible quantum faces a situation reminiscent of the Collingridge dilemma. In this dilemma, the moment in which discussion on societal risks and benefits can be most impactful is also the time where the least information is available. The flipside of this challenge is an opportunity to build processes for examining the public good of quantum before the trajectory (and potential problems) of the sector become 'locked in.' Recent work in this space has argued that quantum researchers and innovators must work with society to address uncertainties and concerns. By engaging quantum stakeholders and understanding their perspectives on responsibility, this paper seeks to support this proposition and enable further dialogue on responsible development and use of quantum technologies.

**Introduction**

>*Interviewer:* What does responsibility look like for quantum?
>
>*Interviewee:* Yeah, so I mean now you've asked a big question and the short answer is I don't know. I don't think anyone knows… we want to start out with conversations with experts in order to start figuring out the direction to go.

Quantum technologies – or technologies that use the science of quantum physics to improve current or create new capabilities – are the subject of intense public and private investment and media scrutiny. They are in varying stages of development and deployment across the world. The growing movement of quantum from the focus of largely fundamental research to the dominion of national strategies and special purpose acquisitions has prompted calls to "begin paying attention to the ethical, social, legal, and economic implications" [1] of the emerging sector.

The field of responsible innovation (RI) encourages us to explore and imagine different forms of responsibility [2-4] and presents approaches for investigating the societal implications of quantum. In particular, the field asks science and innovation actors from diverse sectors to pursue scientific research and technology development responsibly [5]. In this context, responsibility can include the promotion of beneficial applications, responding to ethical concerns, integrating public values, avoiding harm, and supporting human rights (e.g. see OECD Recommendations for Responsible Innovation in Neurotechnology [6]).

Calls to ensure the responsible development of quantum technologies are complicated by a lack of concrete case studies or real-world examples of irresponsible quantum. Technologies in this field currently exist along a spectrum that stretches from imagined applications of basic science through to early field trials. In this state, quantum technologies are not at the stage of 'plug and play'. As a result, responsible quantum faces a situation reminiscent of the Collingridge dilemma [7]. In this dilemma, the moment in which discussion on societal risks and benefits can be most impactful is also the time where the least information is available. The flipside of this challenge is an opportunity to build processes for examining the public good of quantum before the trajectory (and potential problems) of the sector become 'locked in' [8]. As Coenen et al. [9] have recently argued, now is the time for quantum researchers and innovators to "engage transparently with society to find out how they can resonate with public concerns and address social uncertainties."

This paper focuses on researcher and innovator engagement with responsibility in the context of quantum technologies. Here, I report findings from a series of interviews conducted with people acting within roles along the continuum of basic research through to applied work in industry. I start by introducing the project before discussing key themes from the interviews which define responsibility and locate roles in a 'responsible quantum' sector while identifying potential risks. Following this, I draw on participant perspectives to consider how building a language for responsible quantum might enable productive discussion about the societal aspects of the sector. The research questions that drive this analysis were: (1) how do quantum researchers and innovators conceptualise responsibility?, (2) how do these stakeholders conceptualise the risks posed by quantum technology for the society they will exist within?, and (3) what do they consider to be the next steps for developing quantum technologies in a responsible manner?.

**The context of 'responsible quantum'**
Existing academic works on the societal aspects of quantum technology have reviewed the applications of the sector for Defence and security [10]; considered the early ethical implications of quantum research and technologies [11]; reviewed what – if any – public good framing is used in national strategies for the field [12]; emphasised the need to strategically plan for quantum education and workforce development [13]; and highlighted how academic publications on quantum technologies has so far rarely-engaged with social research agendas, including equity, diversity, and inclusion frameworks [14].

In addition to these research publications, recent World Economic Forum discussions over the ethics of quantum computing have resulted in the publication of early principles to govern the emergence of the technology [15]. Meanwhile, the influence of geopolitical tension over quantum computation is visible in rising discussion over export controls and retaining sovereignty [16]. These concerns – sparked by a technology which does not yet exist in any meaningful or useful form – may have follow on impacts on the another prominent concern facing quantum computation: increasing inequality through the creation of quantum 'haves and have-nots' [17]. Work within the physics community to advance the conversation on ethical development of quantum technologies also includes a 'Quantum Ethics' call to action through a mini-documentary launched by quantum sector publication Quantum Daily, a roadmap for ethical quantum computing [18], this special issue on the quantum revolution, and a focus issue on societal aspects of quantum technologies in Quantum Science and Technology [19].

One of the challenges of developing an approach to 'responsible quantum' is defining who is responsible for leading these conversations and how they might be best equipped to create changes to research and innovation practices [3, 20]. Within the context of the demands of responsible innovation and like fields, there remains a gap in terms of how science and innovator actors can be empowered to 'do' responsible research and technology development. Along these lines, work on how responsible innovation might work for and benefit scientists has called for a language of responsibility that has "traction with [scientists] lived experiences, concerns, and terminologies" and relates to everyday scientific practice [21].

**Research design and methods**
Research in quantum science and technology has been underway in Australia since the 1980s [22]. At present, Australia is also home to an increasing number of start-ups and more established businesses developing software and hardware within the umbrella of quantum technologies. The drive to coordinate research and technology translation in quantum has occurred in parallel in other nations, producing a common language around motivation for investment [12]. In some countries, this activity has raised questions over ethical implications of the technologies the strategies aim to support [1].

This study explored how people working in or affiliated with quantum technology in Australia thought about responsibility and the societal implications of quantum technologies. Interviewees were researchers in academia and industry, entrepreneurs, and other relevant industry stakeholders. The participants for this study were selected based on the sectors identified within the Australian CSIRO Quantum Technology Roadmap [23], which forecast applications for quantum technology in various Australian industries and provided an overview of the researchers and entrepreneurs operating in this space. Over the course of this project, the impetus for discussion on responsible development and use of quantum technologies in Australia has increased with a forthcoming national strategy for the sector [24].

During semi-structured interviews that ran for thirty minutes to an hour, interviewees reflected on a range of topics. These included the public communication of quantum technologies, different sources of technical, investment, and societal risk, and pathways for managing the uncertainties of novel technology development. Twenty interviews were carried out in total over a six-month period. Interviewees were recruited by email. The ethical aspects of this research were approved by the University of Queensland Human Research Ethics committee (Project ID: 2020/HE001841).

*Table 1: Overview of interviewees by sector and role*

|  | **Number of interviewees by sector & role** |
|---|---|
| **Higher education** |  |
| Pure research | 6 |
| Research and translation | 3 |
| Building industry linkages & capacity | 2 |
| **Defence** | 2 |
| **Industry** |  |
| Secure communication | 2 |
| Computing | 1 |
| Artificial intelligence/Machine learning | 1 |
| Sensing & imaging | 2 |
| Entrepreneurship and innovation | 1 |
| **Total** | **20** |

Interviews were semi-structured and conducted using a list of core questions (see Table 2) which were adapted to each interviewee's role and experience with the emerging quantum sector. Transcripts from the interviews were analysed thematically and hand coded using a grounded theory approach. This approach allowed for the identification of shared concepts and consistent ideas around 'responsible quantum' across the interviews. In this paper, interview responses are discussed with a focus on the language of responsibility which is emerging within the quantum sector. This paper discusses the key themes that emerged from the interviews. These themes were identified through a process of coding interview transcripts. The draft analysis was reviewed with research participants to ensure accuracy of interpretation prior to submission.

*Table 2: Generic interview questions, adapted for each interviewee during semi-structured discussion*

| **Interview questions** |
|---|
| Can you tell me about yourself and your role? |
| How do you define quantum technology? |
| How do you define quantum computing? |
| Who are quantum technologies being designed for? |
| What are the risks posed by quantum technologies? |
| What might responsibility look like for the quantum sector? |
| What is missing from this conversation? |

*Limitations*

The interviews conducted in this study focused on the conceptualisation of responsibility and risk by quantum stakeholders in Australia. One limitation of this study is the geographical representation. Interviewees were all located in Australian metropolitan areas with concentrated expertise in quantum science and technology. At this stage, there has not been any significant divergence across different geographies in relation to the rhetoric driving investment in quantum technologies. So, the interviews documented in this paper with Australian researchers and industry stakeholders can contribute to international conversations over how best to manage the future impacts of quantum technology.

Another limitation of this study is the diversity of interviewees. In terms of occupation, each was involved or affiliated with quantum research or innovation. The positionality of these interviewees affects their perspectives on the questions raised in interviews. Additional perspectives from societal stakeholders – including policy makers, end users, and people likely to be affected by adopted of quantum technologies – are needed and should be included in future work. The need for diversity in the quantum workforce was also evident in the interviewees engaged for this project, which underscores the need to involve and centre historically underrepresented groups in science and innovation at all stages [25].

**Characterising responsibility in the quantum technology sector**

> "I have a pretty hard time seeing how anything that I do would directly lead to damaging technologies... That being said, when you don't know what you're looking for, you don't know what you're going to find and don't know what can be done with it, so I guess in that sense it's possible." – Quantum researcher (academic)

The conversations carried out in this project prompted reflections on responsibility. Answers to questions on what responsibility meant in the context of developing quantum technologies varied widely, usually in relation to the central professional role of the interviewee. Quantum researchers with minimal ties to translational work generally perceived responsibility around technology development as something located far from their immediate domain. By comparison, quantum researchers who were more involved in the translation of quantum research into technologies had more concrete concerns around responsibility. Those researchers considered the likely societal implications of those technologies (for instance, the impact of dual-use technologies) and most often made comparisons to the impacts of artificial intelligence and machine learning (AI/ML). The comparisons to AI/ML and existing ethical frameworks or principles also served as shorthand for solving the problem of responsibility posed by quantum. Some researchers, though, were more concerned with whether quantum posed novel ethical questions.

One interviewee – a researcher whose role had shifted to a strong focus on developing a quantum industry in Australia – said: "What's really interesting at least to me at first and I think to a lot of the quantum community is: does quantum introduce new ethical or responsible questions? Or do they just amplify and accelerate the classical ones?" This was a point for reflection raised by multiple interviewees, usually physicists, who were interested in whether the introduction of new technologies prompted novel research questions in terms of ethics and impact. This point was also raised in hypothetical rebuttals to individuals within the quantum community who were considered resistant to the discussion of 'responsible quantum.' Resistance was generally framed by an argument that this discussion was either too early or unnecessary given prior work in other fields, including AI/ML and nuclear physics. An interviewee with positions in industry and academia said:
> "One pushback you often hear is people say… that this is true of all of science, technology and engineering, like we don't need to elevate somehow quantum computing above computing or… nuclear physics. Isn't this just the same?"

This interviewee argued that quantum technologies would likely be different due to the nature of the science. They said: "I think it's different because we don't yet understand... what is enabled by this yet to be, you know, leveraged new physics." The urgency to understand this difference was heighted by the financial and intellectual investment poured into the field at a "huge rate" which created a "risk of us getting ahead of ourselves" as different parties compete to build new quantum technologies.

Industry-based stakeholders were more interested in the challenges of scaling up quantum technologies. These conversations tended to centre on the specific outcomes that quantum technologies might enable. Quantum was simultaneously a risk and a buzzword; one that was not yet driving buying decisions, but which offered a potential step change for technology. These stakeholders were more familiar with everyday governance and accountability structures. They referenced the need for regulation that supported translation of quantum technologies while mitigating potential harms. While comparing quantum with the parallel discussions around ethical AI, one industry stakeholder described an approach for considering the limitations of new technologies and identifying areas for concern. They said industry needed to consider how:

> "You address those concerns and think about governance and make sure you've got the right structures on your board and … in your workplace to make sure that you're not doing harm."

As described by the quote above, at least some of these conversations on responsible quantum are motivated by market conditions, or the aspects that might affect the success of businesses within the quantum technology sector.

Interest in more applied approaches to governance – as opposed to the discovery of novel ethical dilemmas – is reflected in events occurring internationally. In parallel to these interviews, uncertainty around the development and governance of future quantum technologies prompted industry-led discussions on ethics and risk at the World Economic Forum. An industry-based interviewee said that these discussions were prompted by "people really having a greater visibility into the fact that quantum technologies are emerging… indeed quantum technologies have entered the planning horizon of many large enterprises." This interviewee described the topics discussed at the forum as: technical risk, risk to shared infrastructure, issues of explainability and verifiability, and national security. Aspects of these topics emerged later in other interviews in relation to risks related to quantum technology.

Defence-focused interviewees said that societal implications – or ethics around deploying quantum technologies – were not yet a central topic of discussion. Instead, they spoke of the atmosphere of quantum technology development nationally, describing it as a "bit of an arms race." The gains promised by quantum for these stakeholders were centred around sovereign capability and new possibilities for Defence capabilities. While comparing autonomous weapons and Defence-related applications of quantum, one interviewee remarked: "I don't see the same level of ethical challenge at this point in time in quantum technologies, because they are providing capabilities essentially that we already have." Instead, these considerations were the subject to a 'watching brief' as technological efforts continued apace.

**Locating responsibility**
> "Part of the challenge is there's nothing, we don't have standards yet, and our customers… tend to be largely driven by what they need to meet" – Industry stakeholder (research and development, executive)

Interviewees located responsibility differently depending on their roles and experience in the emerging sector. Researchers reflected on questions of responsibility in relation to individual conduct, choices, and morals. Designing and conducting research was exciting, but sometimes became problematic with pressures and incentives around success in academia encouraging high levels of hype. One researcher said that in "the world of science and academic, hype and optimism is rewarded and caution and pessimism are penalised." Hype, here, was seen as overwhelming optimism for the field without tempering realism. It influenced promises to funders, the content of journal articles, and the training of postgraduate students.

Researchers who had moved into start-ups also voiced concern over whether jobs existed for the number of postgraduate students trained in quantum science. Describing the push to train a 'next generation' of quantum physicists and engineers, one interviewee said: "[but] that means we've actually got to make it real, right? There's got to be a job for all of these people that we're training." They were also focused on barriers facing those looking to translate their work into a commercial product with a lack of support from research institutions around "translation and the communication with industry."

Industry professionals felt that the promise of the potential sector was "a bit opaque" at present. Uncertainty surrounding quantum technologies meant that the pathway for responsibly implementing these technologies was unclear. One interviewee said that while quantum had potential to solve some important and complex problems, the sector needed to understand the limitations of the technologies at hand. Drawing on lessons from AI/ML, they asked: "Is there any bias in it? It is going to be inherently good for one group of people and not for another?" Other factors influencing interviewee perspectives on this included: lack of standards, concern over the eventual form of regulation, and the absence of ethical boundaries and frameworks.

The form of accountability for "people at the top" of larger enterprises was part of this discussion. In large corporations, "accountability stops and ends at the board" which introduced the need to equip those people with the appropriate knowledge and tools for managing novel technologies. This need circled back to conversations around frameworks and ethics for responsible quantum. One industry-based interviewee said:
> "What do we do if we own the computing power that can now lay bare everyone's secrets? Obviously, it's immoral and unethical to apply it in that way, but using that extreme example, where are our frameworks and guidance to say: OK, these things are now possible, but we're not going to do it."

The role imagined for policy amid these discussions was a combination of providing better support for the burgeoning industry and guidance for implementation. Policy makers needed to not set policy "in a vacuum" but instead providing "guidance and assistance" to "help people live up to what society's expectations are around privacy and security and progress and fairness." This needed to be an "ongoing conversation" that dealt with the export of the technology but also created clear expectations around sovereignty and exclusivity. This kind of reassurance would help drive investment decisions. Standards were vital in this space and processes such as the National Institute of Standards and Technology's call for development of a "new set of quantum safe encryption algorithms" played a significant role in shifting sector-based conversations.

These roles are summarised in Figure 1, below. This diagram uses responses from participants in relation to roles for individuals and organisations in 'responsible quantum' using an adapted triple helix innovation model. The model – represented originally in Kimatu (2016) – usually depicts interactions within the innovation ecosystem between universities, industry, and government. Here, these interlocking relationships are used to reflect on how interviewees imagined different perspectives or roles in relationship to assuming responsibility for quantum technologies.

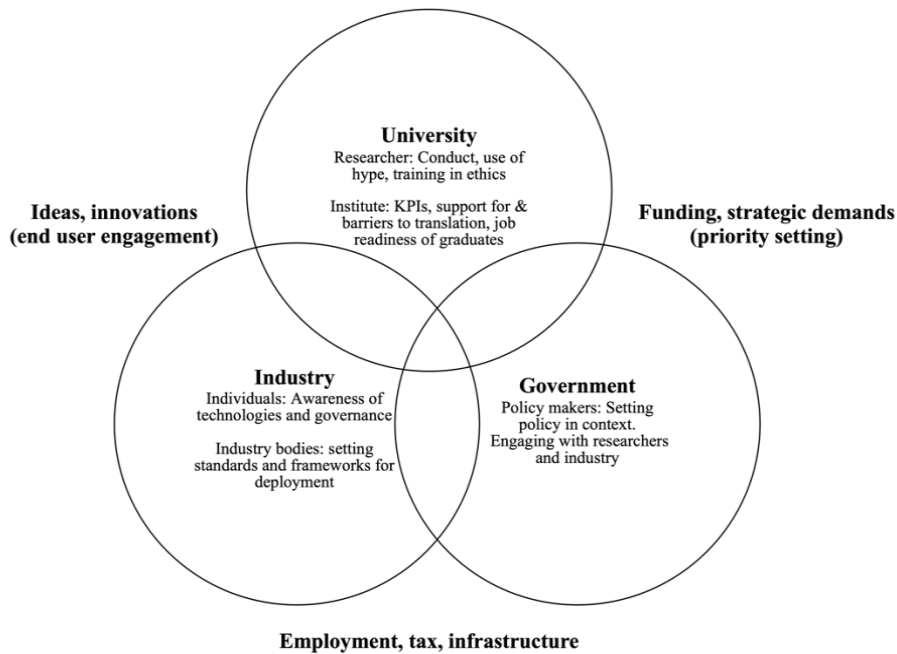

*Figure 1: Summarising roles imagined for actors and institutions in 'responsible quantum'; adaption of triple helix diagram from Kimatu [26]*

**Risks posed by quantum technology**

> "For me, a watching brief [is] where this is all heading. And if it starts to head down somewhere where there might be ethical concerns. Then at that point just holding it and saying OK, should we be doing this?" – Defence stakeholder (manager)

More than thirty themes around risk emerged from the interviews. For the purposes of this article, these have been reduced to nine themes referenced by at least four participants. The dominant concerns cited by interviewees are outlined below and summarised in Table 3. With the specific impacts of quantum technologies still the subject of speculation, discussion over the risks posed by quantum technologies was generally high-level. As a field or discipline, quantum science and technology was perceived as fast moving. This was itself a source of risk, largely in terms of hype or "perception and what perception drives." The challenge of the technology "coming alive in its own right" was cited as a type of technical risk where nations and organisations might face the challenge of becoming 'quantum safe' earlier than anticipated. The costs of choosing to develop quantum technologies (e.g. environmental costs, see work by McKay [27]) were juxtaposed to the costs of not supporting the sector.

Interviewees echoed concerns raised by organisations such as the World Economic Forum regarding sovereignty and inequity in the distribution of and access to quantum technologies. As a Defence stakeholder described it as important that they build "sovereign capability… [because] we want to do as best we can to keep that within Australia so we're not reliant on buying stuff from overseas." Other interviewees supported this view that retaining some level of exclusive capability was a necessity. However, there was high levels of uncertainty around what form this might take. One researcher said: "there needs to be an ongoing conversation with government, not just about the export of the technology, but also retaining some capability, sovereignty, exclusivity." There were also concerns over the implications of retaining this exclusive use. An industry-based interviewee agreed that quantum technologies were likely to "confer strategic and commercial advantage." The question then was whether they "become controlled technologies? Will they be restricted for national purposes? And, could this in turn then lead to a set of quantum haves and quantum have nots?" As a researcher described, "a lot of countries are not investing in quantum research. What you will have is some countries with very developed technology and they will be the gatekeepers to these technologies." This state meant that countries without access to quantum would inevitably have to "trust [in] these early players" but this trust would put them in a disadvantaged, vulnerable state.

The exclusionary nature of quantum technologies was further complicated by the barriers of cost and skilled workers. Interviewees suggested that one avenue for democratising access to quantum resources was supporting cloud-based infrastructure, such as the IBM Q experience for quantum computing. However, this potential solution raises questions of data disclosures and research ownership. The drive for sovereignty was not only likely to impact countries and individuals in the future. Rising concerns over exclusive capabilities was impacting researchers in the present. For example, conflicting ideas about retaining sovereign use is has recently influenced discussions over quantum computing research in Europe [28]. Tension over research on emerging technology is also likely to influence government restrictions on foreign research collaboration in Australia [29]. The high level of expectations surrounding quantum was perceived to be an additional problem in this space. For instance, one researcher considered the hype of quantum computing to be so influential that any government making a claim to have such technology would have a weapon in the sense that "it would be a symbol of technology, of technological prowess, it would be something that would instil fear." They argued that "at some point we have to think about how this kind of technology could change the balance of power in certain situations", in some cases before the capability might even be technically available.

The prevalence of hype around quantum technologies was a key concern with eight interviewees discussing exaggerated promises and the 'selling' promissory culture of the sector. Hype around timelines for development and uses cases for yet unrealised technology risked causing a 'quantum winter.'[1] In relation to funding and industry interest, one interviewee said: "I can see potential for the undelivered promise of quantum to have a longer term impact on both quantum sciences and other emerging technologies." A researcher working on a start-up company described the impact of quantum hype as a complicating factor for talking about their products. They said, "I would say it's difficult talking about quantum stuff in general nowadays [because of previous hype] … we had to be really careful to not promise too much, especially now that we're into the commercial world." Later, the same interviewee noted that there needs to be a shift from "these amazing promises" that might be remotely possible to "understanding where the problems are and understand that they are immediate problems that can be solved, you know, today." For another researcher with an emerging industry presence, responsibility meant "not lying… not misrepresenting what the opportunity is" for investors and other audiences. Perhaps related to this state of hype was the risk of technical failure, the question of "are things actually going to work?" Part of the challenge here was the shifting culture between academia and industry with an accompanying need to move away from open questions and ambitious promises encouraged within academia.

---

[1] A term adapted from 'AI winter' in which the field saw a lack of funding and social support, this is generally assumed to be caused by hype and unmet expectations.

Layered onto – and perhaps influencing – the issue of hype was the two-pronged opacity problem posed by quantum technologies. The science underlying the technologies meant that explainability and verifiability (in an ethical AI sense) will be impossible. One industry stakeholder described this as:

> "So, with other kinds of technology, you can clearly say: OK, you know, I put this in here. Here's the chain of events that will occur and here I get the output. With quantum… those are still areas where work needs to be done for people to gain confidence in the technology"

This opacity issue was compounded by the difficulty of communicating about the science itself. Responsibility for addressing this situation was generally located with the individual researchers or start-up. One interviewee described this in terms of the challenge faced by the deep tech industry in general where new products were often integrated within complex systems. "End users [often] don't understand enough about quantum or whatever the tool might be to understand what impact it's going to have" which meant the company needed to be able to convey "how the systems that they're currently using… can be augmented by quantum." The challenge facing quantum as an emerging industry then was engaging with other sectors and possible end users to understand how quantum technologies might have impact.

The incentives for success in academia and industry were considered a complicating factor and a risk. One researcher said that the two avenues for success in the field were, for academics, "publish[ing] technical things with big names" or, in a corporate career, "mak[ing] money for the company rather than discover foundational things that are good for the field." From their perspective, these measures of success were problematic when it came to how to incentivise "ethical work." Another researcher described choices in academia as:

> "How do I go about doing my research is the first thing. How do I talk about it and sell it so that I responsibly sell it and not blow it up to absolute impossibility? How do I go about responsibly informing people of where it leads and what it does?"

Few, if any, of the risks identified so far for quantum technologies were distinct from other discussions over the implications of novel technologies. This was recognised in several of the interviews with interviewees identifying the fields – predominantly artificial intelligence and machine learning – they felt quantum could learn from. Although, one interviewee who was involved in ethical AI said:

> "I think there's quantum specific ethics to consider… [some elements that] potentially could be backwards influencing. I know at the moment I'm saying quantum should look a lot at AI ethics, but I think AI ethics would really benefit [from quantum]."

Another (quantum) researcher said:

> "We're just beginning to understand, I think, what kind of power being able to use entanglement for technological purpose brings. It's just very hard to see that future… I guess the risk is that indeed there are applications here that we haven't yet discovered."

*Table 3: Risks posed by and for second quantum revolution*

| Categories | Risks | Who is responsible? |
|---|---|---|
| **Security and access** | Sovereignty and quantum technologies | Government; policy makers; Defence |

|  | National security, dual-purpose technologies | Defence |
|---|---|---|
|  | Inequity in the distribution of and access to quantum technologies (related to sovereignty) | All stakeholders |
| **Clarity around boundaries** | Lack of regulation, standards | Government; policy makers; industry bodies |
| **Communication** | Hype and a possible quantum winter | Researchers; industry |
|  | Opacity problem; lack of discussion around specific quantum ethics | Researchers |
| **Technological** | Technical failure | Researchers; industry |
| **Incentives for success** | Responsible conduct in research and training | Researchers and research institutions; research funders |
| **Unknown future risks** | Unknown/unforeseen impacts | All stakeholders |

**Where to from here? Building awareness of 'responsible quantum'**
> "I think with any nascent or emerging technology. It's important that as a society we do think through and ask the questions… and don't just blindly forge ahead just because we know how to do it." – Industry stakeholder (research and development, executive)

How do we move forward with the notion of 'responsible quantum'? Half of the interview participants said that we needed to start a conversation on responsibility and quantum technology. Another two made implicit references to starting discussions, though one of these participants was referred more to industry engagement rather than conversation on societal implications more broadly. One researcher referenced communication in the sense of demystifying quantum technology and preparing people. They said: "it would be good to better communicate to people…that this stuff is coming" because "it would be better if people were more prepared for it." Building on this notion of preparation, one interviewee reflected on the need to increase awareness of "responsible innovation or responsible use, equitable use" among incoming quantum science graduates and workers. They said "awareness, if you like, is the absolute minimum that … we need to do."

Awareness and preparation were dominant themes. Interviewees suggested that while researchers and students should build their understanding around societal and ethical concerns, policy makers and regulators should build their own language around quantum. An interviewee said: "you have to get the regulators used to speaking in terms of quantum. This is that it's not always this magical strange word… we have to have these conversations about things like equitable access." An additional space for better communication was researcher engagement with industry. Here, one academic argued that "we just need a better conversation between the quantum physicists and the people who actually do this for a living" to ensure that quantum technologies have a useful impact in various sectors.

On the theme of societal impact, some interviewees also identified a need to talk about the common good or public good of quantum. One researcher with a growing industry presence said that:
> "There are some important things for the public good as well that aren't necessarily money making, but there for the public good. It would be good to be able to articulate that context in a way that isn't a copy and paste from the last grant."

Building on this theme in reference to previous discussions on sovereignty, one industry stakeholder asked: "What should we be doing to ensure that the greatest global good, if you will, is derived from these technologies, so it's not stymied… by national security considerations?"

The flipside of this conversation was captured by one Defence-focused researcher who outlined the conundrum facing researchers who "think that they're working on civilian projects that have societal good ends, but that those technologies could be on-sold or developed for [different] purposes" than those that the "researchers were hoping to achieve." The action items that came through these discussions were two-fold. The first was this general call for further conversation between stakeholders and disciplines around how to develop quantum technologies responsibly. The second was a call to shift quantum research culture through more interdisciplinary training for students and "downward pressure" from funders, conferences, and journals who could motivate consideration of societal impacts. These ideas were primarily drawn from developments within the AI/ML community, which include increasing expectation that researchers will review the societal and ethical aspects of their work at different stages of research.

These questions of how and when to begin a conversation on the societal aspects of quantum have emerged across multiple forums. For instance, within a small workshop held on the emerging topic of quantum artificial intelligence (quantum AI) in Noosa, Australia. The workshop was multidisciplinary and drew on fields that included physics, philosophy, and social science. Conversations around the overarching theme of the conference were chiefly focused on how to define the landscape of quantum AI and determine research directions. Despite this, attendees were keen to reflect on the possible societal aspects of this potential technology. Drawing on the example of 'classical' (current) AI/ML approaches, they argued that it was important to think early about these technological advances and identify the 'uniquely quantum' challenges posed. According to these researchers, the first step for such discussion was cross-pollination or involving researchers from different disciplines to help embed questions within everyday workplace culture and create a more conscious dialogue with external stakeholders.

These discussions and the content of the interviews analysed in this paper share a set of common concepts with calls for conversations on quantum technology. For one, these conversations are imagined as nuanced with regards to hype (less sensationalised) and open to wider expertise on societal or ethical elements. Interviewees framed these conversations as driven by the desire to identify and, perhaps, redesign pressures that impact responsible conduct. These concepts in addition to the headline risks identified by participants suggest a way forward in terms of building a shared language around 'responsible quantum' which crosses over the boundaries of academia, industry, government, and wider civil society.

**Conclusion and next steps**
This project aimed to understand how quantum stakeholders imagined the responsible development and deployment of quantum technologies. The interviews uncovered the topics prioritised by different stakeholders and drew out the pre-existing factors that will influence 'responsible quantum' in the future. Further work is needed to develop case studies of different quantum technologies that identify benefits and risks. These cases could be used to invite wider (public) perspectives of the sector and add in the civil society component of the innovation helix (see Kimatu [26] for an example of this).

In this study, interviewees positioned quantum as an essential, positive development for Australian industry. The continued development and deployment of quantum technologies was optimistically imagined. Risks and responsibility were imagined in relation to experiences of pressures and possibilities within specific institutional settings. Ethical elements of quantum were chiefly discussed in terms of novelty. In other word, academic researchers in quantum chiefly wanted to discover and address ethical concerns for quantum that were fundamentally different to ethical questions posed by other, existing technologies. While a critical analysis of these underlying assumptions was beyond the scope of this study, they are still worth noting given these assumptions will continue to shape the possibilities for quantum research and technology going forward.

Calls for responsibility in quantum and further afield require additional work from science and innovation actors. As Glerup, Davies [21] note, understanding the context that these actors work within as well as the structures and tensions that enable and complicate responsible innovation may help promote responsibility that has traction in their everyday worlds. Based on the findings here it seems that in quantum there is an acknowledged need for processes and tools that support responsible development and use of these emerging technologies. This article goes some way towards identifying the concepts that future work will need to address.

**Acknowledgements:** With thanks to interviewees who gave up their time to participate in this study as well as thanks to F. Medvecky and referees for thoughtful comments and discussion.

**Funding:** The author undertook this research and prepared the manuscript while funded by the Australian Research Council Centre of Excellence for Engineered Quantum Systems (EQUS, CE170100009).